\documentclass[a4paper,11pt]{article}
\pdfoutput=1 

\usepackage{jheppub} 
\usepackage{etoolbox}
    \makeatletter
    \patchcmd{\maketitle}{\@fpheader}{}{}{}
    \makeatother                     

\usepackage[T1]{fontenc} 
\usepackage{slashed}

\newcommand{\be}{\begin{equation}}
\newcommand{\ee}{\end{equation}}
\newcommand{\bea}{\begin{eqnarray}}
\newcommand{\eea}{\end{eqnarray}}

\title{\boldmath Cosmology of  Linear Higgs-Sigma Models with Conformal Invariance}

\preprint{}

\author[]{Hyun Min Lee}
\author[]{and Adriana G. Menkara}

\affiliation[]{Department of Physics, Chung-Ang University, Seoul 06974, Korea}

\emailAdd{hminlee@cau.ac.kr }
\emailAdd{amenkara@cau.ac.kr }

\abstract{
We consider general linear Higgs-sigma models  as ultra-violet completions of the Higgs inflation. We introduce general higher curvature terms beyond Einstein gravity and recast them into a class of linear Higgs-sigma models in the scalar-dual formulation where conformal symmetry is manifest. Integrating out the sigma field in this class of linear sigma models, we obtain the same Higgs inflation Lagrangian of non-linear sigma model type in the effective theory. 
We show that the successful inflation for sigma field singles out the sigma-field potential derived from the $R^2$ term and  the tracker solution for dark energy at late times can be realized for the $R^{p+1}$ term with $-1<p<0$. 
We also discuss the implications of Higgs-sigma interactions for the inflation and the vacuum stability in the Standard Model.
}

\begin{document} 
\maketitle
\flushbottom

\section{Introduction}
\label{sec:intro}

The origins of cosmic inflation in the early Universe and cosmological acceleration at late times are among the most important questions in modern cosmology to date. The former period with large vacuum energy solves various problems in Standard Big Bang cosmology such as horizon problem, homogeneity, isotropy, etc, and it seeds the large-scale structure of the Universe. On the other hand, the latter period with dark energy would determine the fate of the Universe in the future.

Higgs inflation \cite{Higgsinf} has drawn attention due to the minimal possibility that the Higgs boson in the Standard Model is the inflaton with a non-minimal coupling to gravity.  Starobinsky inflation \cite{starobinsky} is also the economic extension of General Relativity with an $R^2$ term, whose predictions for inflationary observables are in perfect agreement with Planck data \cite{planck2018}, similarly as in Higgs inflation.

In the original version of Higgs inflation, however, there is an unitarity problem due to the large non-minimal coupling, associated with the would-be Goldstone bosons in the Higgs doublet \cite{unitarity}. Thus, there is a need of introducing new degrees of freedom to restore the unitarity in Higgs inflation up to the Planck scale, so there have been a number of suggestions in the context of linear sigma model analogues for chiral perturbation theory in QCD \cite{sigma,general,higgsportal,hmlee,espinosa}. Recently, it has been shown that the $R^2$ term can provide an ultra-violet (UV) completion for Higgs inflation \cite{R2inflation,scpark} and linear sigma models for Higgs inflation were identified in the basis with conformal invariance \cite{ema,ema2}.

In this article, we consider general linear Higgs-sigma models as UV completions of the Higgs inflation. In this framework, the linear Higgs-sigma models are regarded as being basis-independent under conformal transformations, and conformal symmetry is explicitly broken in the Einstein term and the scalar potential. 
We introduce general higher curvature terms beyond Einstein gravity and derive the corresponding linear Higgs-sigma models by taking the dual-scalar formulation and identifying it as a sigma field. Then, we analyze the inflation dynamics in the linear Higgs-sigma models and compare the results to those in the literature where the conformal  invariance was not manifest. 
We also obtain the new results for the tracker solution for dark energy from the sigma-field potential derived from general higher curvature terms and compare the time-varying equation of state for dark energy to the observed data. 

The paper is organized as follows.
We introduce the Higgs inflation as non-linear sigma models in the basis with conformal invariance and review the strategy for that in detail. We make the non-linear sigma models in linear forms for the Starobinsky model as well as the analogues of general higher curvature terms. Then, we discuss the inflation in linear sigma models motivated by the UV completion into the Starobinsky model. Next we show the implications of the sigma-field potential for dark energy in the linear sigma models derived from general higher curvature terms. Finally, conclusions are drawn. There are two appendices showing the relation to sigma models of different form and identifying the Higgs interactions in going from conformal frame to Einstein frame.

\section{Higgs inflation as non-linear sigma models}

We consider the Lagrangian for Higgs inflation with non-minimal coupling $\xi$, as follows,
\bea
{\cal L}=\sqrt{-{\hat g}} \bigg[-\frac{1}{2}(1+\xi {\hat\phi}^{2}_i)  {\hat R}+\frac{1}{2} g^{\mu\nu} \partial_\mu {\hat\phi}_i \partial_\nu {\hat\phi}_i -\frac{\lambda}{4} ({\hat\phi}^2_i)^2 \bigg]. \label{Higgsinf}
\eea
In the original UV completion for Higgs inflation \cite{sigma}, the Higgs kinetic terms are identified as non-linear sigma models in Einstein frame and they are linearized by the introduction of a real-singlet sigma field.
In this work, taking the basis with conformal invariance \cite{ema,ema2},  we first review the details on how to recast the Higgs inflation in non-linear sigma model and introduce a sigma field after that.

\subsection{The conformally invariant Lagrangian}

First, we introduce an unphysical scalar degree of freedom in the metric, the conformal mode $\varphi$, in the following,
\bea
{\hat g}_{\mu\nu}=e^{2\varphi} {g}_{\mu\nu}.
\eea
Then, using
\bea
{\hat R}&=&e^{-2\varphi} {R}- 6 e^{-3\varphi} \Box e^\varphi, \\
\sqrt{-{\hat g}}&=&e^{4\varphi}\,\sqrt{-{g}},
\eea 
we can rewrite the Lagrangian (\ref{Higgsinf}) as
\bea
{\cal L}&=&\sqrt{-{g}}\, e^{4\varphi} \bigg[ -\frac{1}{2} e^{-2\varphi}(1+\xi {\hat\phi}^2_i)  {R}+ 3 (1+\xi{\hat\phi}^2_i)  e^{-3\varphi} \Box e^\varphi \nonumber  +\frac{1}{2} e^{-2\varphi} (\partial_\mu{\hat\phi}_i)^2 -\frac{\lambda}{4} ({\hat\phi}^2_i)^2 \bigg] \nonumber \\
&=&\sqrt{-{ g}}\, \bigg[ -\frac{1}{2} e^{2\varphi}(1+\xi {\hat\phi}^2_i)  {R}+ 3 (1+\xi{\hat\phi}^2_i)  e^{\varphi} \Box e^\varphi \nonumber  +\frac{1}{2} e^{2\varphi} (\partial_\mu{\hat\phi}_i)^2 -\frac{\lambda}{4} \,e^{4\varphi}({\hat\phi}^2_i)^2 \bigg].
\eea
Next, making the field redefinitions by ${\phi}_i=e^{\varphi} {\hat\phi}_i$ and ${\Phi}=\sqrt{6}\,e^\varphi$, we get
\bea
{\cal L}&=&\sqrt{-{\hat g}} \bigg[  -\frac{1}{2} \Big(\frac{1}{6} {\Phi}^2+\xi {\phi}^2_i \Big) {R} -\frac{1}{2} (\partial_\mu{\Phi})^2 +3\xi\phi^2_i\, \frac{\Box {\Phi}}{{\Phi}} +\frac{1}{2}{\Phi}^2 \Big[\partial_\mu ({\Phi}^{-1}{\phi}_i)\Big]^2-\frac{\lambda}{4}({\phi}^2_i)^2  \bigg] \nonumber \\
&=&\sqrt{-{\hat g}} \bigg[  -\frac{1}{2} \Big(\frac{1}{6} {\Phi}^2+\xi {\phi}^2_i \Big) {R} -\frac{1}{2} (\partial_\mu{\Phi})^2+\frac{1}{2} (\partial_\mu{\phi}_i)^2 +3\Big(\xi+\frac{1}{6} \Big)\phi^2_i\,\frac{\Box {\Phi}}{{\Phi}}-\frac{\lambda}{4}({\phi}^2_i)^2 \bigg]\label{Higgsinf2}
\eea
where we made integrations by parts.

The next step is to redefine ${\Phi}=\phi+\sigma$ and get
\bea
\frac{\Box {\Phi}}{{\Phi}}=(\phi+\sigma)^{-1} (\Box \phi + \Box \sigma).
\eea
Then, we take the following combination of the conformal mode and the Higgs fields,
\bea
\frac{1}{6} {\Phi}^2 +\xi {\phi}^2_i =\frac{1}{6} \phi^2-\frac{1}{6} {\phi}^2_i -\frac{1}{6} \sigma^2,
\eea
or
\bea
\frac{1}{6} (\phi+\sigma)^2 +\xi {\phi}^2_i =\frac{1}{6} \phi^2-\frac{1}{6} {\phi}^2_i -\frac{1}{6} \sigma^2.
\eea
As a result, we get the above equation to solve for $\sigma$ as follows,
\bea
\sigma=\frac{1}{2} \bigg( \sqrt{\phi^2-12\Big(\xi+\frac{1}{6}\Big) {\phi}^2_i}-\phi\bigg). \label{constraint0}
\eea
This is nothing but the constraint equation for the $\sigma$ field.
Then, with $\sigma={\Phi}-\phi$, the relation with the redefined conformal mode $\phi$ is given by
\bea
{\Phi} =\frac{1}{2} \bigg( \sqrt{\phi^2-12\Big(\xi+\frac{1}{6}\big) {\phi}^2_i}+\phi\bigg).
\eea

Finally, from
\bea
\bigg[\frac{1}{2}{\Phi}^2+ 3\Big(\xi+\frac{1}{6} \Big){\phi}^2_i\bigg]\,\frac{\Box {\Phi}}{{\Phi}}&=&\frac{1}{2} (\phi^2-\sigma^2) \,\frac{\Box {\Phi}}{{\Phi}} \nonumber \\
&=& \frac{1}{2} (\phi-\sigma) (\Box \phi + \Box \sigma) \nonumber \\
&=&-\frac{1}{2} (\partial_\mu\phi)^2 + \frac{1}{2} (\partial_\mu  \sigma)^2,
\eea
where we made integration by parts again, we can rewrite eq.~(\ref{Higgsinf}) in the final form,
\bea
{\cal L}= \sqrt{-{g}} \bigg[  -\frac{1}{2} \Big(\frac{1}{6} {\phi}^2-\frac{1}{6} {\phi}^2_i-\frac{1}{6}\sigma^2 \Big) { R} -\frac{1}{2} (\partial_\mu{\phi})^2+\frac{1}{2} (\partial_\mu{\phi}_i)^2+ \frac{1}{2} (\partial_\mu \sigma)^2  -\frac{\lambda}{4}({\phi}^2_i)^2 \bigg]. \label{conformalL}
\eea
Therefore, in the above final form of the Lagrangian, the non-minimal coupling for the Higgs fields in the Higgs inflation is moved into the non-canonical kinetic term in $(\partial_\mu\sigma)^2$ with the constraint, eq.~(\ref{constraint0}). If the $\sigma$ field is promoted to a dynamical field as will be discussed in the next section, all the scalar fields couple conformally to gravity, so the above Lagrangian is invariant under the local conformal transformations.

\subsection{Gauge-fixed Lagrangian}

Now we fix the gauge for the conformal mode to $\phi=\sqrt{6}$ to determine the Planck mass.
Then, we obtain eq.~(\ref{conformalL}) in the conformally invariant form except the Einstein term,
\bea
{\cal L}= \sqrt{-{g}} \bigg[  -\frac{1}{2} \Big(1-\frac{1}{6} {\phi}^2_i-\frac{1}{6}\sigma^2 \Big) { R} +\frac{1}{2} (\partial_\mu{\phi}_i)^2+ \frac{1}{2} (\partial_\mu \sigma)^2  -\frac{\lambda}{4}({\phi}^2_i)^2 \bigg]
\eea
with
\bea
f(\sigma,\phi_i)\equiv \bigg(\sigma+\frac{\sqrt{6}}{2}\bigg)^2 +3\Big(\xi+\frac{1}{6} \Big){\phi}^2_i-\frac{3}{2}=0. \label{constraint}
\eea
Here, we rewrote eq.~(\ref{constraint0}) in a more illuminating form, $(2\sigma+\phi)^2 +12\Big(\xi+\frac{1}{6} \Big){\phi}^2_i=\phi^2$, which is then recasted into eq.~(\ref{constraint}).  The constraint equation (\ref{constraint}) defines the vacuum manifold for the would-be linear sigma model with the $\sigma$ field.
Therefore, we can regard the Higgs inflation as non-linear sigma models with the constraint equation for the sigma field given in eq.~(\ref{constraint}).

\subsection{Promoting the constraint to a dynamical field}

We can introduce the constraint equation  (\ref{constraint}) as a Lagrange multiplier $\kappa(x)$ in the Lagrangian, in the following form,
\bea
\Delta{\cal L}= -\sqrt{-{g}}\,  \frac{\kappa(x)}{4}\Big(f(\sigma,\phi_i)\Big)^2.
\eea
Then, we can recover the Lagrangian in eq.~(\ref{conformalL}) by using the equation of motion for $\kappa(x)$. 
But, if the $\sigma$ field is promoted to a dynamical field as in linear sigma models, then the Lagrange multiplier becomes a coupling parameter, and we obtain a UV model with the sigma field included, with the following Lagrangian in the fixed gauge,
\bea
{\cal L}&=& \sqrt{-{g}} \bigg\{  -\frac{1}{2} \Big(1-\frac{1}{6} {\phi}^2_i-\frac{1}{6}\sigma^2 \Big) {R} +\frac{1}{2} (\partial_\mu{\phi}_i)^2+ \frac{1}{2} (\partial_\mu \sigma)^2  \nonumber \\
&&-\frac{\lambda}{4}({\phi}^2_i)^2-\frac{\kappa}{4}\left[\bigg(\sigma+\frac{\sqrt{6}}{2}\bigg)^2 +3\Big(\xi+\frac{1}{6} \Big){\phi}^2_i- \frac{3}{2}\right]^2 \bigg\}. \label{sigmamodels}
\eea
In this case, as far as the non-minimal couplings, the kinetic terms and the Higgs quartic coupling are concerned, the local conformal symmetry is respected. But, the conformal symmetry is broken explicitly by both the Planck mass and the $\sigma$ potential. Nonetheless, we dub the chosen metric frame ``the conformal frame'', because the scalar fields couple non-minimally to gravity in a conformally invariant way.

From the sigma model Lagrangian in eq.~(\ref{sigmamodels}), it is obvious that there is no unitarity violation below the Planck scale, and the theory is perturbative, as far as the quartic couplings for the sigma field and the Higgs quartic coupling are small, i.e.
\bea
\kappa\lesssim 1,\qquad \lambda+9\kappa \Big(\xi+\frac{1}{6} \Big)^2\lesssim 1, \qquad 6\kappa \Big(\xi+\frac{1}{6} \Big)\lesssim 1. \label{pert}
\eea

We note that the sigma-field potential in eq.~(\ref{sigmamodels}) can be generalized to any function of the constraint $f(\sigma,\phi_i,)$, as far as there exists a local minimum with $f(\sigma,\phi_i)=0$. As will be shown in the next section, the general sigma-field potential can be obtained from higher curvature terms in Higgs inflation.

\section{General linear sigma models from higher curvature terms}

In this section, we derive the linear sigma model from the Starobinsky model with $R^2$ term. Then,  linear sigma models are generalized to the case with a general curvature expansion. We give explicit examples for the cases where one or two terms are dominant in the curvature expansion.

\subsection{Starobinsky model as a linear sigma model}

Suppose that an $R^2$ term is added in the Lagrangian for Higgs inflation in eq.~(\ref{Higgsinf}), as follows,
\bea
{\cal L}_{R2}=\sqrt{-{\hat g}} \bigg[-\frac{1}{2}(1+\xi {\hat\phi}^2_i)  {\hat R}+\frac{1}{2} g^{\mu\nu} \partial_\mu {\hat\phi}_i \partial_\nu {\hat\phi}_i -\frac{\lambda}{4} ({\hat\phi}^2_i)^2 +\alpha {\hat R}^2\bigg]. \label{r2}
\eea
Then, in the conformal frame for the dual scalar field (scalaron) corresponding to the $R^2$ term,
the quartic coupling for the sigma field in eq.~(\ref{sigmamodels}) can be identified as
\bea
\kappa = \frac{1}{36\alpha}. \label{r2-quartic}
\eea
We present the details on the derivation of the above result below.

First, we introduce a dual scalar field $\hat\chi$ for the $R^2$ term and rewrite the Lagrangian (\ref{r2}) as
\bea
\frac{{\cal L}_{R2}}{\sqrt{-{\hat g}}}= -\frac{1}{2} {\hat R} (1+\xi {\hat\phi}^2_i + 4\alpha {\hat\chi}) -\alpha{\hat\chi}^2  + \frac{1}{2} (\partial_\mu{\hat\phi}_i)^2 -\frac{\lambda}{4} ({\hat\phi}^2_i)^2  \label{r2-dual}
\eea
Then, performing the Weyl transformation by ${\hat g}_{\mu\nu}=\Omega^{-2} g_{\mu\nu}$ and the field redefinitions, ${\hat\phi}_i=\Omega \phi_i$ and ${\hat\chi}=\Omega^2\chi$, 
we get 
\bea
\frac{{\cal L}_{R2}}{\sqrt{-g}}&=&\Omega^{-4}\bigg[ \frac{1}{2}\Omega^2 \Big(-R+6(\partial_\mu\ln \Omega)^2-6\partial^2\ln\Omega\Big) (1+\xi\Omega^2\phi^2_i+4\alpha\Omega^2\chi ) \nonumber \\
&&-\alpha\Omega^4 \chi^2 +\frac{1}{2} \Omega^2 \Big( \partial_\mu(\Omega\phi_i)\Big)^2 -\frac{\lambda}{4} \Omega^4\phi^4_i \bigg] \nonumber \\
&=&-\frac{1}{2} (\Omega^{-2} +\xi\phi^2_i +4\alpha\chi) R +3\Omega^{-2} \Big((\partial_\mu\ln\Omega)^2-\partial^2\ln\Omega \Big) (1+\xi \Omega^2\phi^2_i+4\alpha\Omega^2\chi) \nonumber \\
&&-\alpha\chi^2 + \frac{1}{2} \Omega^{-2} (\Omega\partial_\mu\phi_i+\partial_\mu\Omega\,\phi_i)^2 - \frac{\lambda}{4} \phi^4_i . \label{r2a}
\eea
Here, we note that the following derivative terms can be rewritten as
\bea
\Omega^{-2} (\Omega\partial_\mu\phi_i+\partial_\mu\Omega\,\phi_i)^2
&=&(\partial_\mu\phi_i)^2 +(\partial_\mu\ln\Omega)\, \phi_i\partial^\mu\phi_i +  (\partial_\mu\ln\Omega)^2 \phi^2_i \nonumber \\
&=& (\partial_\mu\phi_i)^2 +\phi^2_i \Big((\partial_\mu\ln\Omega)^2-\partial^2\ln\Omega \Big), 
\eea
up to a total derivative term.
Thus, we can recast eq.~(\ref{r2a}) into
\bea
\frac{{\cal L}_{R2}}{\sqrt{-g}}&=&-\frac{1}{2} (\Omega^{-2} +\xi\phi^2_i +4\alpha\chi) R +3\Big((\partial_\mu\ln\Omega)^2-\partial^2\ln\Omega \Big) \bigg(\Omega^{-2} +\Big(\xi+\frac{1}{6}\Big)\phi^2_i+4\alpha\chi\bigg) \nonumber \\
&&-\alpha\chi^2 + \frac{1}{2}  (\partial_\mu\phi_i)^2 - \frac{\lambda}{4} \phi^4_i. \label{r2b}
\eea

Now we choose the conformal factor by
\bea
\Omega^{-2}= \Big(1+\frac{\sigma}{\sqrt{6}} \Big)^2 \label{r2-conf}
\eea
with the following constraint for the sigma field $\sigma$,
\bea
\Omega^{-2} +\xi\phi^2_i +4\alpha\chi= 1-\frac{1}{6} \phi^2_i -\frac{1}{6}\sigma^2. \label{r2-conf2} 
\eea
As a result, from
\bea
\Omega^{-2} +\Big(\xi+\frac{1}{6}\Big)\phi^2_i+4\alpha\chi= 1-\frac{1}{6}\sigma^2,
\eea
we can rewrite the part of the kinetic term as
\bea
3\Big((\partial_\mu\ln\Omega)^2-\partial^2\ln\Omega \Big) \bigg(\Omega^{-2} +\Big(\xi+\frac{1}{6}\Big)\phi^2_i+4\alpha\chi\bigg)
&=&\frac{3}{\sqrt{6}}\Big(1-\frac{1}{6}\sigma^2\Big) \Omega\partial^2\sigma \nonumber \\
&=&-\frac{1}{2} \sigma \partial^2\sigma,
\eea
up to a total derivative term.
Finally, from eq.~(\ref{r2b}) with the above result, we obtain the Lagrangian in the conformally invariant form, as follows,
\bea
\frac{{\cal L}_{R2}}{\sqrt{-g}}=-\frac{1}{2} R\Big(1-\frac{1}{6}\phi^2_i -\frac{1}{6} \sigma^2 \Big)+\frac{1}{2} (\partial_\mu\sigma)^2 +\frac{1}{2}(\partial_\mu\phi_i)^2 -\alpha \chi^2 -   \frac{\lambda}{4} \phi^4_i,
\eea
with
\bea
\chi=\frac{1}{4\alpha} \bigg[\frac{1}{2}-\frac{1}{3} \Big(\sigma+\frac{\sqrt{6}}{2}  \Big)^2-\Big(\xi+\frac{1}{6}\Big)\phi^2_i \bigg]. 
\eea
Here, we have used eqs.~(\ref{r2-conf}) and (\ref{r2-conf2}) to express $\chi$ in terms of the other fields.
As a result, the above sigma field Lagrangian coincides with the linear sigma model derived from Higgs inflation in eq.~(\ref{sigmamodels}). So,  from
\bea
U(\sigma,\phi_i) = \alpha \chi^2 = \frac{1}{16\alpha}\, \bigg[\frac{1}{2}-\frac{1}{3} \Big(\sigma+\frac{\sqrt{6}}{2}  \Big)^2-\Big(\xi+\frac{1}{6}\Big)\phi^2_i \bigg]^2, \label{r2-pot}
\eea
we can identify the quartic coupling for the sigma field as in eq.~(\ref{r2-quartic}).
In this case, the perturbativity conditions in eq.~(\ref{pert}) become
\bea
\frac{1}{36\alpha} \lesssim 1, \qquad \lambda+ \frac{1}{4\alpha} \Big(\xi+\frac{1}{6} \Big)^2\lesssim 1,  \qquad \frac{1}{4\alpha}\,  \Big(\xi+\frac{1}{6} \Big)\lesssim 1, \label{pertcond}
\eea
which are consistent with the unitary conditions in the mixed Higgs-$R^2$ inflation \cite{R2inflation}.

\subsection{General linear sigma models}

Now we consider the general linear sigma models by taking the extension of the Higgs inflation with $R^{k+1}$ curvature term with $k>0$, as follows,
\bea
{\cal L}_{\rm gen}=\sqrt{-{\hat g}} \bigg[-\frac{1}{2}(1+\xi {\hat\phi}^2_i)  {\hat R}+\frac{1}{2} g^{\mu\nu} \partial_\mu {\hat\phi}_i \partial_\nu {\hat\phi}_i -\frac{\lambda}{4} ({\hat\phi}^2_i)^2 +\sum_k\frac{2(-1)^{k+1}\alpha_k}{k+1}\, {\hat R}^{k+1}\bigg]\label{rn}
\eea
with $\alpha_k$ being coupling parameters in the curvature expansion.
Then, introducing a dual scalar field $\hat\chi_k$ for each $R^{k+1}$ term, we obtain the dual-scalar Lagrangian as
\bea
\frac{{\cal L}_{\rm gen}}{\sqrt{-{\hat g}}}= -\frac{1}{2} {\hat R} \Big(1+\xi {\hat\phi}^2_i + \sum_k 4\alpha_k {\hat\chi}_k\Big) -\sum_k 2\Big(\frac{k}{k+1}\Big)\,\alpha_k\,{\hat\chi}^{\frac{k+1}{k}}_k  + \frac{1}{2} (\partial_\mu{\hat\phi}_i)^2 -\frac{\lambda}{4} ({\hat\phi}^2_i)^2.  \label{rk-dual}
\eea

Following the similar steps with the field redefinitions, ${\hat g}_{\mu\nu}=\Omega^{-2} g_{\mu\nu}$, ${\hat\phi}_i=\Omega \phi_i$ and ${\hat\chi}_k=\Omega^{2}\chi_k$, with $\Omega^{-2}=\big(1+\frac{\sigma}{\sqrt{6}}\big)^2$, as in the previous subsection, we find the corresponding Lagrangian in the conformally invariant form, as follows,
\bea
\frac{{\cal L}_{\rm gen}}{\sqrt{-g}}&=&-\frac{1}{2} R\Big(1-\frac{1}{6}\phi^2_i -\frac{1}{6} \sigma^2 \Big)+\frac{1}{2} (\partial_\mu\sigma)^2 +\frac{1}{2}(\partial_\mu\phi_i)^2 \nonumber \\
&& -\sum_k \Omega^{-2+\frac{2}{k}}\Big(\frac{2k}{k+1}\Big)\,\alpha_k\,{\chi}^{1+\frac{1}{k}}_k   -   \frac{\lambda}{4} \phi^4_i \label{genL}
\eea
where the constraint equation for $\chi_k$ is given by
\bea
\sum_k 4\alpha_k\chi_k=\frac{1}{2}-\frac{1}{3} \Big(\sigma+\frac{\sqrt{6}}{2} \Big)^2-\Big(\xi+\frac{1}{6}\Big)\phi^2_i. \label{constraint-gen}
\eea
Then, we can introduce the above constraint equation as a Lagrange multiplier $y(x)$,
\bea
\frac{\Delta{\cal L}_{\rm gen}}{\sqrt{-g}}= y(x)\cdot \left[\sum_k 4\alpha_k\chi_k- \frac{1}{2}+\frac{1}{3} \Big(\sigma+\frac{\sqrt{6}}{2} \Big)^2+\Big(\xi+\frac{1}{6}\Big)\phi^2_i\right].
\eea
As a consequence, varying the full Lagrangian with respect to $\chi_k$ and $y$, we obtain the dual scalar fields in terms of the Lagrange multiplier $y$ as
\bea
\chi_k = 2^k\Omega^{2k-2} \,y^k \label{duals}
\eea
where $y$ satisfies the following equation,
\bea
\sum_k 4\alpha_k \,2^k \Omega^{2k-2} \,y^k=\frac{1}{2}-\frac{1}{3} \Big(\sigma+\frac{\sqrt{6}}{2} \Big)^2-\Big(\xi+\frac{1}{6}\Big) \phi^2_i. \label{lagmul}
\eea
For $k=1,2,\cdots, N$,  the Lagrange multiplier $y$ is the solution to the $N$-th order algebraic equation, which is a function of $\sigma$ and $\phi_i$. 

Finally, from eq.~(\ref{genL}) with eq.~(\ref{duals}), we obtain the sigma field potential in terms of the the Lagrange multiplier $y$ as
\bea
U(\sigma,\phi_i)&=& \sum_k \Omega^{-2+\frac{2}{k}}\Big(\frac{2k}{k+1}\Big)\,\alpha_k\,{\chi}^{1+\frac{1}{k}}_k \nonumber \\
 &=&\sum_k \Big(\frac{2^{k+2} k}{k+1} \Big) \alpha_k (\Omega(\sigma))^{2k-2} (y(\sigma,\phi_i))^{k+1}.  \label{genpotg}
\eea
Then, the decoupling condition for the sigma field, $\frac{\partial U}{\partial\sigma}=0$, gives rise to
\bea
0= \frac{\partial U}{\partial\sigma}&=& \sum_k  2^{k+2} k  \alpha_k (\Omega(\sigma))^{2k-2} (y(\sigma,\phi_i))^k\Big(\frac{\partial y}{\partial\sigma} \Big)  \nonumber \\
&&-\frac{1}{\sqrt{6}} \sum_k \Big(\frac{2^{k+2} k}{k+1} \Big) \alpha_k(2k-2) (\Omega(\sigma))^{2k-1} (y(\sigma,\phi_i))^{k+1}.
\eea
Therefore, as far as $k\geq 1$, there always exists an extremum for $y=0$, which corresponds to the constraint equation for the non-linear sigma model in eq.~(\ref{constraint}). As a result, after integrating out the sigma field with $y=0$ for the general higher curvature terms, we get the same non-linear sigma model for Higgs inflation below the mass of the sigma field as for Starobinsky model. 

For consistent UV complete models, we need similar perturbativity conditions as in eq.~(\ref{pertcond}), which restrict the form of general higher curvature terms and the non-minimal coupling for the Higgs field. We discuss the details on the UV complete models in the following examples.

\subsubsection{Example 1: A single higher curvature term}

If only the $R^{p+1}$ term with the coefficient $\alpha_p$ is nonzero and $\alpha_k=0$ for $k\neq p$, the constraint equation in eq.~$(\ref{constraint-gen})$ becomes
\bea
\chi_p=\frac{1}{4\alpha_p} \bigg[\frac{1}{2}-\frac{1}{3} \Big(\sigma+\frac{\sqrt{6}}{2} \Big)^2-\Big(\xi+\frac{1}{6}\Big)\phi^2_i \bigg].
\eea
Then, we obtain the generalized scalar potential for the sigma field in the conformal frame, as follows,
\bea
U(\sigma,\phi_i)&=&\Omega^{-2+\frac{2}{p}}\Big(\frac{2p}{p+1}\Big)\, \alpha_p \chi^{1+\frac{1}{p}}_p \nonumber \\
&=& \frac{1}{3}\cdot 2^{-1-\frac{2}{p}}\Big(\frac{p}{p+1}\Big)\,\Big(\frac{1}{3\alpha_p}\Big)^{\frac{1}{p}}\Big(1+\frac{\sigma}{\sqrt{6}}\Big)^{2\big(1-\frac{1}{p}\big)} \nonumber \\
&&\times \left[ \frac{3}{2}-\bigg(\sigma+\frac{\sqrt{6}}{2}\bigg)^2 -3\Big(\xi+\frac{1}{6} \Big){\phi}^2_i\right]^{1+\frac{1}{p}} \nonumber \\
&\equiv& \frac{1}{4} \kappa_n (\sigma+\sqrt{6})^{4(1-n)}\left[-\sigma(\sigma+\sqrt{6})-3\Big(\xi+\frac{1}{6} \Big){\phi}^2_i\right]^{2n}  \label{genpot}
\eea
with $n=\frac{1}{2}\big(1+\frac{1}{p}\big)$. 
We note that the overall constant $\kappa_n$ in the scalar potential is proportional to $(\alpha_p)^{-1/|p|}$ for $p>0$ (or $n>\frac{1}{2}$) but to $(\alpha_p)^{1/|p|}$  for $p<0$ (or $n<\frac{1}{2}$). Thus, as will be discussed in the later sections, a large value of $\alpha_p$ is favored for inflation with $p>0$ whereas a small value of $\alpha_p$ for dark energy with $p<0$.

First, we note that  the sigma model potential becomes singular at $U=0$ for non-integer $\frac{1}{p}$ (or 2n) and $ \alpha_p>0$, so the field range is bounded as follows,
\bea
\sigma(\sigma+\sqrt{6}) +3\Big(\xi+\frac{1}{6} \Big){\phi}^2_i<0,
\eea
in addition to $\phi^2_i + \sigma^2<6$ for the positive effective Planck mass from eq.~(\ref{genL}). For non-integer $\frac{1}{p}$ (or $2n$) and $\alpha_p$, the potential becomes negative, so it is not appropriate for inflation.
On the other hand, for integer $\frac{1}{p}$ (or $2n$), the potential is bounded from below, only  if $\frac{1}{p}$ (or $2n$) is odd.

Imposing $\frac{\partial U}{\partial \sigma}=0$ for the effective potential in eq.~(\ref{genpot}), we can identify the vacuum manifold for the sigma model: for $n>\frac{1}{2}$ (or $p>0$),
\bea
\sigma(\sigma+\sqrt{6} ) +3\Big(\xi+\frac{1}{6} \Big){\phi}^2_i=0 \, ;
\eea 
for $n<\frac{1}{2}$ (or $p<0$),
\bea
(\sigma+\sqrt{6}) \Big(\sigma+\frac{1}{2}n\sqrt{6} \Big) +3(1-n) \Big(\xi+\frac{1}{6} \Big){\phi}^2_i=0.
\eea
In the former case, we can get the same vacuum structure with $\langle\sigma\rangle=\langle\phi_i\rangle=0$ (for ignoring the electroweak scale) as in Starobinsky model,  so we can recover the Higgs inflation in the effective theory after integrating out the sigma field. On the other hand, in the latter case, the vacuum structure is totally different from the one for Starobinsky model, namely, $\langle\sigma\rangle=-\sqrt{6}$ or $\sigma=-\frac{1}{2}n\sqrt{6}$ for $\langle\phi_i\rangle=0$.  
Therefore, we can regard the former case with  $n>\frac{1}{2}$ (or $p>0$) as being candidates for UV complete models for the Higgs inflation because they have the same vacuum structure as in the Starobinsky model. 
On the other hand, the latter case with  $n<\frac{1}{2}$ (or $p<0$) is a distinct class of models, which are disconnected from the Starobinsky model, and it is subject to a further stability check. In the later section, we will make a separate discussion on this class of models for dark energy. 

We now consider the perturbativity constraints on the effective couplings derived from higher curvature terms. As we discussed for the dual-scalar theory of Starobinsky model, the pertubativity conditions are dominated by the Higgs self-interactions in eq.~(\ref{pertcond}). Assuming that the sigma field gets a nonzero VEV, $\langle\sigma\rangle={\cal O}(1)$, in the presence of an extra higher curvature term, we can expand the sigma potential in eq.~(\ref{genpot}) for the Higgs self-interactions, as follows,
\bea
{\cal L}_{\rm eff} =-\frac{1}{4}\kappa_n(-1)^{2n} \sigma^{2n}(\sigma+\sqrt{6})^{2(2-n)}\sum_{l=0}^\infty\left(\begin{array}{c} 2n \\ l \end{array}\right) \frac{3^l(\xi+\frac{1}{6})^l(\phi^{2}_i)^l}{\sigma^l(\sigma+\sqrt{6})^l} \label{series}
\eea
where the series is infinite for non-integer $2n$. 
Therefore, the perturbativity on the extra Higgs quartic coupling gives rise to $\kappa_n (\xi+\frac{1}{6})^2\lesssim 1$, while higher order terms for Higgs self-interactions with $l>2$ are suppressed by the Planck scale, under the conditions,  $\kappa_n (\xi+\frac{1}{6})^l\lesssim 1$, which are  stronger than the one for the Higgs quartic coupling. We note that for integer $2n$, the series in eq.~(\ref{series}) terminates at a finite order with maximum power of $(\phi^2_i)^{2n}$, so the strongest perturbative bound in this case becomes $\kappa_n (\xi+\frac{1}{6})^{2n}\lesssim 1$.

\subsubsection{Example 2: $R^2+ R^3$}

If only $R^2$ and $R^3$ terms are nonzero, eq.~(\ref{lagmul}) becomes
\bea
8(\alpha_1 y + 2\alpha_2 \Omega^2 y^2)=\frac{1}{2}-\frac{1}{3} \Big(\sigma+\frac{\sqrt{6}}{2} \Big)^2-\Big(\xi+\frac{1}{6}\Big) \phi^2_i\equiv -\frac{1}{3} f(\sigma,\phi_i). \label{r2r3}
\eea
Then, solving the above quadratic equation for $y$, we get the Lagrange multiplier for $y>0$ as
\bea
y=\frac{1}{4\alpha_2} \,\Omega^{-2} \Big(-\alpha_1+\sqrt{\alpha^2_1-\frac{1}{3} \alpha_2\Omega^2 f } \Big). \label{lagmul-sol}
\eea
Here, we note that for the real solution for $y$ to exist, the field space is bounded to $f(\sigma,\phi_i)<3\alpha^2_1 \Omega^{-2}/\alpha_2$. 
Therefore, from eq.~(\ref{genpotg}), we obtain the dual scalar potential as
\bea
U(\sigma,\phi_i) 
&=& 4\alpha_1 y^2 + \frac{32}{3} \alpha_2 \Omega^2 y^3 \nonumber \\
&=&\frac{1}{12\alpha^2_2}\,\Omega^{-4} \bigg[ 2\Big(\alpha^2_1-\frac{1}{3}\alpha_2\Omega^2 f\Big) \sqrt{\alpha^2_1-\frac{1}{3} \alpha_2\Omega^2 f} -\alpha_1 (2\alpha^2_1-\alpha_2\Omega^2 f) \bigg].  \label{r2r3pot}
\eea
Here, we used eq,~(\ref{lagmul-sol}) in the last equality.

For a sizable value of $\alpha_2$, we can expand the sigma potential in eq.~(\ref{r2r3pot}), as follows,
\bea
U(\sigma,\phi_i)&=&  \frac{\alpha^3_1}{6\alpha^2_2}\,\Omega^{-4}\bigg[ \Big(1-\frac{\alpha_2}{3\alpha^2_1}\Omega^2 f  \Big)\sum_{l=0}^\infty \left(\begin{array}{c} \frac{1}{2} \\ l \end{array}\right) \Big(- \frac{\alpha_2}{3\alpha^2_1}\Omega^2 f\Big)^l-\Big(1- \frac{\alpha_2}{2\alpha^2_1}\Omega^2 f \Big)\bigg]  \nonumber \\
&=&   \frac{\alpha^3_1}{12\alpha^2_2} \sum_{l=2}^\infty \frac{\sqrt{\pi}}{\Gamma\big(\frac{3}{2}-l\big)(l-1)!} \Big(\frac{1}{l} -\frac{2}{3}(-1)^l\Big)    \Big( \frac{\alpha_2}{3\alpha^2_1}\Big)^l\Omega^{2l-4} \nonumber \\
&&\quad\times \bigg[\sigma(\sigma+\sqrt{6})+3\Big(\xi+\frac{1}{6}\Big) \phi^2_i\bigg]^l \label{r2r3series}
\eea
where we used eq.~(\ref{r2r3}) in the last line. 
As a result, we find that the vacuum manifold is given by $f=0$ or $\sigma(\sigma+\sqrt{6})+3\Big(\xi+\frac{1}{6}\Big) \phi^2_i=0$ as for Starobinsky model. Moreover, from the Higgs self-interactions at a given $l$ in eq.~(\ref{r2r3series}), we can impose the perturbativity bounds on $(\phi^2_i)^l$ with $l\geq 2$ as $ \frac{1}{\alpha_1}\big(\alpha_2/\alpha^2_1\big)^{l-2}  (\xi+\frac{1}{6})^l\lesssim 1$.  For instance, for $l=2$, we obtain the perturbative bound, $(\xi+\frac{1}{6})^2/\alpha_1\lesssim 1$, which is the same as in eq.~(\ref{pertcond}). But, for $l=m+2$ with $m\geq 1$, we need extra suppression factors for perturbativity by $\big(\alpha_2/\alpha^2_1\big)^m (\xi+\frac{1}{6})^m\lesssim \big(\alpha_2/\alpha^2_1\big)^m \alpha^{m/2}_1\lesssim 1$. In this case, we can regard the derived sigma models coming from $R^2+R^3$ as being the UV completion of the Higgs inflation.

In the limit of a vanishing $\alpha_2$, we can recover the result for Starobinsky model in the previous section, 
$U(\sigma,\phi_i)\approx \frac{1}{144\alpha_1} \, f^2$, which coincides with eq.~(\ref{r2-pot}). Then, the same perturbative constraints as in eq.~(\ref{pertcond}) apply on $\alpha\to\alpha_1$ and $\xi$.

\section{Inflation in linear sigma models}

We discuss the inflationary dynamics and model predictions in linear sigma models that are derived from the $R^2$ term.  Conformal invariance becomes manifest in our approach.

We first choose the unitary gauge for the SM Higgs field such that $\phi_i=(0,0,0,h)^T$.
Then, considering the potential for the sigma field  in eq.~(\ref{genpot}) from the $R^{2}$ term derived in the conformal frame, we consider the full Lagrangian, 
\bea
{\cal L}= \sqrt{-{g}} \bigg\{  -\frac{1}{2} \Big(1-\frac{1}{6} h^2-\frac{1}{6}\sigma^2 \Big) {R} +\frac{1}{2} (\partial_\mu h)^2+ \frac{1}{2} (\partial_\mu \sigma)^2 -\frac{\lambda}{4} h^4-U(\sigma, h) \bigg\} \label{sigmamodels2}
\eea
where the sigma field potential becomes
\bea
U(\sigma,h)=  \frac{\kappa_1}{4}\,\left[\sigma(\sigma+\sqrt{6})+3\Big(\xi+\frac{1}{6} \Big)h^2\right]^{2}. \label{sigmapot}
\eea

Making a Weyl rescaling of the metric by $g_{\mu\nu}=g_{E,\mu\nu}/\Omega^{\prime 2}$ with $\Omega^{\prime 2}=1-\frac{1}{6} h^2-\frac{1}{6}\sigma^2$, from eq.~(\ref{sigmamodels2}), we get the Einstein frame Lagrangian as
\bea
{\cal L}_E&=&  \sqrt{-{g_E}} \bigg\{  -\frac{1}{2}\,R(g_E)+\frac{3}{4\Omega^{\prime 4}} (\partial_\mu\Omega^{\prime 2})^2+\frac{1}{2\Omega^{\prime 2}} (\partial_\mu h)^2+ \frac{1}{2\Omega^{\prime 2}} (\partial_\mu \sigma)^2 -V(\sigma, h)\bigg\}  \nonumber \\
&=& \sqrt{-{g_E}} \bigg\{  -\frac{1}{2}\,R(g_E)+\frac{1}{2\Omega^{\prime 4}} \bigg[\Big( 1-\frac{1}{6}\sigma^2\Big)(\partial_\mu h)^2 +\Big( 1-\frac{1}{6} h^2\Big)(\partial_\mu \sigma)^2+\frac{1}{3} h\,\sigma\, \partial_\mu h \partial^\mu\sigma   \bigg]  \nonumber \\
&&\quad-V(\sigma,h) \bigg\} 
\label{general-einstein}
\eea
where the Einstein frame potential is given by
\bea
V(\sigma,h)&=&\frac{1}{\Omega^{\prime 4}}\,\bigg( \frac{\lambda}{4} h^4+U(\sigma, h) \bigg) \nonumber \\
&=&\frac{1}{\big(1-\frac{1}{6}h^2 -\frac{1}{6}\sigma^2\big)^2} \bigg[\frac{1}{4} \kappa_1 \bigg(\sigma(\sigma+\sqrt{6})+3\Big(\xi+\frac{1}{6} \Big)h^2\bigg)^{2}+\frac{1}{4}\lambda h^4 \bigg]. \label{totalpot}
\eea
Here, the perturbativity sets the limit on the effective running Higgs quartic coupling,
\bea
\lambda_{\rm eff}=\lambda+ 9 \kappa_1 \Big(\xi+\frac{1}{6} \Big)^2\lesssim {\cal O}(1). \label{lameff}
\eea
We remark that after integrating out the sigma field, the effective Higgs quartic coupling is given by $\lambda$, which is smaller than the effective running coupling $\lambda_{\rm eff}$, due to the tree-level threshold correction \cite{vsb}. Therefore, we can solve the vacuum instability problem in the SM for the appropriate choices of $\kappa_1$ and $\xi$.

\subsection{Effective inflaton potential}

In order to obtain the effective inflaton potential for the sigma field, we first integrate the Higgs field by taking the minimization of the total scalar potential in Einstein frame in eq.~(\ref{totalpot}), $\frac{\partial V}{\partial h}=0$. The minimum with $h=0$ is stable as far as the effective Higgs quartic coupling in eq.~(\ref{lameff}) is positive and the effective Higgs mass is large enough. Moreover, we also get a nontrivial condition for $h$, as follows,
\bea
h^2=\frac{\kappa_1\sigma (\sigma+\sqrt{6}) \big(\sigma-3\big(\xi+\frac{1}{6}\big)(\sigma-\sqrt{6}) \big)}{\lambda (\sigma-\sqrt{6})-3\kappa_1\big(\xi+\frac{1}{6}\big)\big(\sigma-3\big(\xi+\frac{1}{6}\big)(\sigma-\sqrt{6}) \big) }. \label{hmin}
\eea 
Here, we note that $h^2$ goes to zero in the limit of $\sigma\to -\sqrt{6}$ during inflation, so we can ignore the kinetic term for $h$ and the kinetic mixing term in eq.~(\ref{general-einstein}).
Then, the kinetic term for the sigma field is approximately the same as the one in pure sigma-field inflation in the previous subsection. 

Now plugging the condition (\ref{hmin}) back to the total scalar potential in eq.~(\ref{totalpot}), we find the effective inflaton Lagrangian, as follows,
\bea
\frac{{\cal L}_{\rm eff}}{\sqrt{-{g_E}} } =  -\frac{1}{2}\,R(g_E)+ \frac{ (\partial_\mu \sigma)^2 }{2(1-\sigma^2/6)^2}- V_{\rm eff}(\sigma) 
\eea
with
\bea
V_{\rm eff}(\sigma)= 9\lambda\,\kappa_1\sigma^2 \bigg[\lambda (\sigma-\sqrt{6})^2+\kappa_1 \bigg(\sigma-3\Big(\xi+\frac{1}{6}\Big)(\sigma-\sqrt{6}) \bigg)^2  \bigg]^{-1}.
\eea
Therefore, in terms of the canonical inflaton field  $\chi$ related to the sigma field by
\bea
\sigma=-\sqrt{6} \tanh \Big(\frac{\chi}{\sqrt{6}}\Big), \label{inflaton}
\eea
 the effective inflaton potential becomes
\bea
V_{\rm eff}(\chi) = \frac{9\kappa_1}{4} \Big(1-e^{-2\chi/\sqrt{6}} \Big)^2 \left[1+\frac{\kappa_1}{4\lambda} \Big(6\xi+e^{-2\chi/\sqrt{6}} \Big)^2 \right]^{-1}. \label{effpot}
\eea

Here, we note that from the minimization condition in eq.~(\ref{hmin}), the Higgs field  during inflation is given in terms of the canonical inflaton field in eq.~(\ref{inflaton}) for $\chi\gg 1$ by
\bea
h^2\simeq \frac{72\kappa_1 \xi}{2\lambda+3\kappa_1 \xi(1+6\xi)}\, \cdot e^{-2\chi/\sqrt{6}}.
\eea
Then, for $ \lambda\sim \kappa_1\xi^2\lesssim 1$ and $\xi\gg 1$, we can approximate $h^2\lesssim \frac{36\kappa_1\xi}{\lambda}\, e^{-2\chi/\sqrt{6}}\ll 6-\sigma^2\simeq 24\,  e^{-2\chi/\sqrt{6}}$. Therefore, the Higgs field contributions to the kinetic terms in eq.~(\ref{general-einstein}) can be neglected for inflation as argued previously. 

In this case, we can recover the pure sigma-field inflation for $9\kappa_1 \xi^2\ll \lambda$ and the Higgs inflation for $9\kappa_1 \xi^2\gg \lambda$, as follows,
\bea
V_{\rm eff}(\chi) \approx \left\{ \begin{array}{c} \frac{9\kappa_1}{4} \Big(1-e^{-2\chi/\sqrt{6}} \Big)^2, \quad 9\kappa_1 \xi^2\ll \lambda,  \\  \frac{\lambda}{4\xi^2} \Big(1-e^{-2\chi/\sqrt{6}} \Big)^2,\quad 9\kappa_1 \xi^2\gg \lambda. \end{array}\right. 
\eea
In general, during inflation for $\chi\gg 1$,  the inflaton vacuum energy  is approximately given by
\bea
V_0=  \frac{9\kappa_1\lambda}{4 (\lambda+9\kappa_1 \xi^2)}.
\eea
Thus, the inflaton vacuum energy depends on Higgs and sigma quartic couplings as well as the non-minimal coupling for the Higgs field. The results are consistent with Ref.~\cite{R2inflation} where conformal invariance was not manifest.

\subsection{Inflationary predictions}

For the sigma-Higgs inflation with eq.~(\ref{effpot}), we get the slow-roll parameters for $\chi\gg 1$, as follows,
\bea
\epsilon &=&  \frac{1}{2}\bigg(\frac{1}{V_{\rm eff}}\frac{dV_{\rm eff}}{d\chi} \bigg)^2= \frac{1}{3} \frac{(2\lambda+3\kappa_1\xi(1+6\xi))^2}{(\lambda+9\kappa_1\xi^2)^2}\,e^{-4\chi/\sqrt{6}} , \\
\eta &=&\frac{1}{V_{\rm eff}}\frac{d^2V_{\rm eff}}{d\chi^2}  =-\frac{2}{3} \, \cdot \frac{2\lambda+3\kappa_1\xi(1+6\xi)}{\lambda+9\kappa_1\xi^2}\, e^{-2\chi/\sqrt{6}}  \nonumber \\
&&\qquad\qquad\quad\,\,+ \frac{2\kappa_1}{3}\, \cdot\frac{(-\lambda+12\lambda\xi+18\kappa_1\xi^2(1+6\xi))}{(\lambda+9\kappa_1\xi^2)^2}\, e^{-4\chi/\sqrt{6}}.
\eea 
On the other hand, the number of efoldings is given by
\bea
N&=& \int_{\chi_e}^{\chi_*} \frac{{\rm sgn}(dV_{\rm eff}/d\chi)\,d\chi}{\sqrt{2\epsilon}} \nonumber \\
&=& \frac{3}{2}\,\cdot\frac{\lambda+9\kappa_1\xi^2}{2\lambda+3\kappa_1\xi(1+6\xi)}\, \Big(e^{2\chi_*/\sqrt{6}} -e^{2\chi_e/\sqrt{6}}\Big)
\eea
where $\chi_*, \chi_e$ are the inflation field values at the horizon exit and the end of inflation, respectively.
Then, for $e^{2\chi_*/\sqrt{6}} \gg e^{2\chi_e/\sqrt{6}}$, we can rewrite the slow-roll parameters at horizon exit in terms of the number of efoldings.  Consequently, we get the spectral index and the tensor-to-scalar ratio in terms the number of efoldings,
\bea
n_s &=& 1-6\epsilon_*+2\eta_* \nonumber \\
&=& 1-\frac{2}{N} -\frac{9}{2N^2} + \frac{3\kappa_1}{N^2}\, \frac{(-\lambda+12\lambda\xi+18\kappa_1\xi^2(1+6\xi))}{(2\lambda+3\kappa_1\xi(1+6\xi))^2},
\eea
and 
\bea
r= 16\epsilon_* = \frac{12}{N^2}. 
\eea
Here, we note that the $1/N^2$ terms are different from the pure sigma inflation or Starobinsky inflation due to the Higgs quartic coupling, but the extra terms are not significant in the perturbative regime with $\lambda\sim \kappa_1\xi^2\lesssim 1$. Thus, the predictions for the spectral index and the tensor-to-scalar ratio are almost the same as in  the pure sigma inflation or Starobinsky inflation \cite{starobinsky,general}. 

Moreover, the CMB normalization constrains the inflation vacuum energy by
\bea
A_s= \frac{1}{24\pi^2}\, \frac{V_0}{\epsilon_*} = 2.1\times 10^{-9},
\eea
resulting in
\bea
\frac{\sqrt{\lambda+9\kappa_1\xi^2}}{\sqrt{\kappa_1\lambda}} = 1.5\times 10^5.
\eea
Then, both the sigma and Higgs quartic couplings contribute to the CMB normalization as in the sigma models of induced gravity type \cite{sigma}. For $9\kappa_1\xi^2\gg \lambda$, we find  $\xi/\sqrt{\lambda}=5\times 10^5$ as in the case for Higgs inflation. But, for $9\kappa_1\xi^2\ll \lambda$, we just get the constraint on the sigma field quartic coupling by $\kappa_1=4\times 10^{-11}$.

\section{Cosmology for general linear sigma models}

When a single $R^{p+1}$ curvature term is added to Einstein gravity with conformal couplings,  the scalar potential for general linear sigma models is given by eq.~(\ref{genpot}). Then, in the unitary gauge for the SM Higgs field, we replace the sigma-field scalar potential in eq.~(\ref{sigmapot}) by
\bea
U(\sigma,h)=  \frac{\kappa_n}{4}\,(\sigma+\sqrt{6})^{4(1-n)}\left[-\sigma(\sigma+\sqrt{6})-3\Big(\xi+\frac{1}{6} \Big)h^2\right]^{2n} \label{sigmapotn}
\eea
with $n=\frac{1}{2} \big(1+\frac{1}{p}\big)$. 
In this section, we obtain the scalar potential for dark energy in Einstein frame and analyze the tracker solutions for dark energy by both analytical and numerical methods.

Our analysis focuses on the specific curvature term of $R^{p+1}$ and its dual description with conformal invariance, but we note that there are more general discussions on modified gravity theories and dark energy in the context of $F(R)$ gravity in the literature \cite{FRgrav,recent}. Nonetheless, in our case,  the non-minimal coupling for the Higgs fields and the higher curvature terms for dark energy can be formulated in a conformally invariant fashion and the conformal breaking effects show up in the sigma model potential. Moreover, we identify the interactions between dark energy and the Higgs fields directly from the sigma potential and discuss the effects of the dark energy on the running Higgs quartic coupling and the vacuum stability in the Standard Model.

\subsection{Dark energy from the sigma field}

We first consider the case where the SM Higgs is decoupled.
Then, setting $h=0$, we can focus on the dynamics of the sigma field only, with the following Einstein-frame Lagrangian,
\bea
\frac{{\cal L}_E}{\sqrt{-{g_E}} } =  -\frac{1}{2}\,R(g_E)+ \frac{ (\partial_\mu \sigma)^2 }{2(1-\sigma^2/6)^2}- V(\sigma) 
\eea
with
\bea
V(\sigma)&=&  \frac{{\kappa}_n}{4(1-\sigma^2/6)^2}\, \cdot (-\sigma)^{2n}(\sigma+\sqrt{6})^{2(2-n)} \nonumber \\
&=& 9{\kappa}_n\,\cdot \frac{(-\sigma)^{2n}(\sigma+\sqrt{6})^{2(1-n)} }{(\sigma-\sqrt{6})^2}.
\eea

Making the sigma kinetic term canonical by the  field redefinition with eq.~(\ref{inflaton}),
we can rewrite the sigma field Lagrangian as
\bea
\frac{{\cal L}_E}{\sqrt{-{g_E}} } =  -\frac{1}{2}\,R(g_E)+ \frac{1}{2} (\partial_\mu\chi)^2 - V(\chi)
\eea
with
\bea
V(\chi)
&=&\frac{9{\kappa}_n}{4^{n}} \,  \cdot e^{-4(1-n)\chi/\sqrt{6}}\Big(1-e^{-2\chi/\sqrt{6}} \Big)^{2n} \nonumber \\
&=&\frac{9\kappa_n}{4^n} \,\cdot e^{-2\big(1-\frac{1}{p}\big)\chi/\sqrt{6}}  \Big(1-e^{-2\chi/\sqrt{6}} \Big)^{1+\frac{1}{p}} \equiv V_0(\chi). \label{pures}
\eea
Then, for $n<1$ (in other words, $p>1$ or $p<0$), the inflaton potential is exponentially suppressed due to the prefactor, so it is quintessence-like for dark energy \cite{quint} rather than for inflation. On the other hand, if $n>1$ (in other words, $p<1$), the overall factor in eq.~(\ref{pures})  becomes exponentially growing for a large $\chi$. 
For $n=1$ (or $p=1$), the sigma field potential in eq.~(\ref{pures}) becomes
\bea
V(\chi) = \frac{9}{4}\,\kappa_1 \Big(1-e^{-2\chi/\sqrt{6}} \Big)^2,
\eea
which corresponds to the $h=0$ case in the previous section and coincides with the one in the Starobinsky model with $\kappa_1 = \frac{1}{36\alpha}$\cite{starobinsky,general}.  This case is appropriate for inflation in the early Universe as discussed in the previous section.

 \begin{figure}
  \begin{center}
    \includegraphics[height=0.45\textwidth]{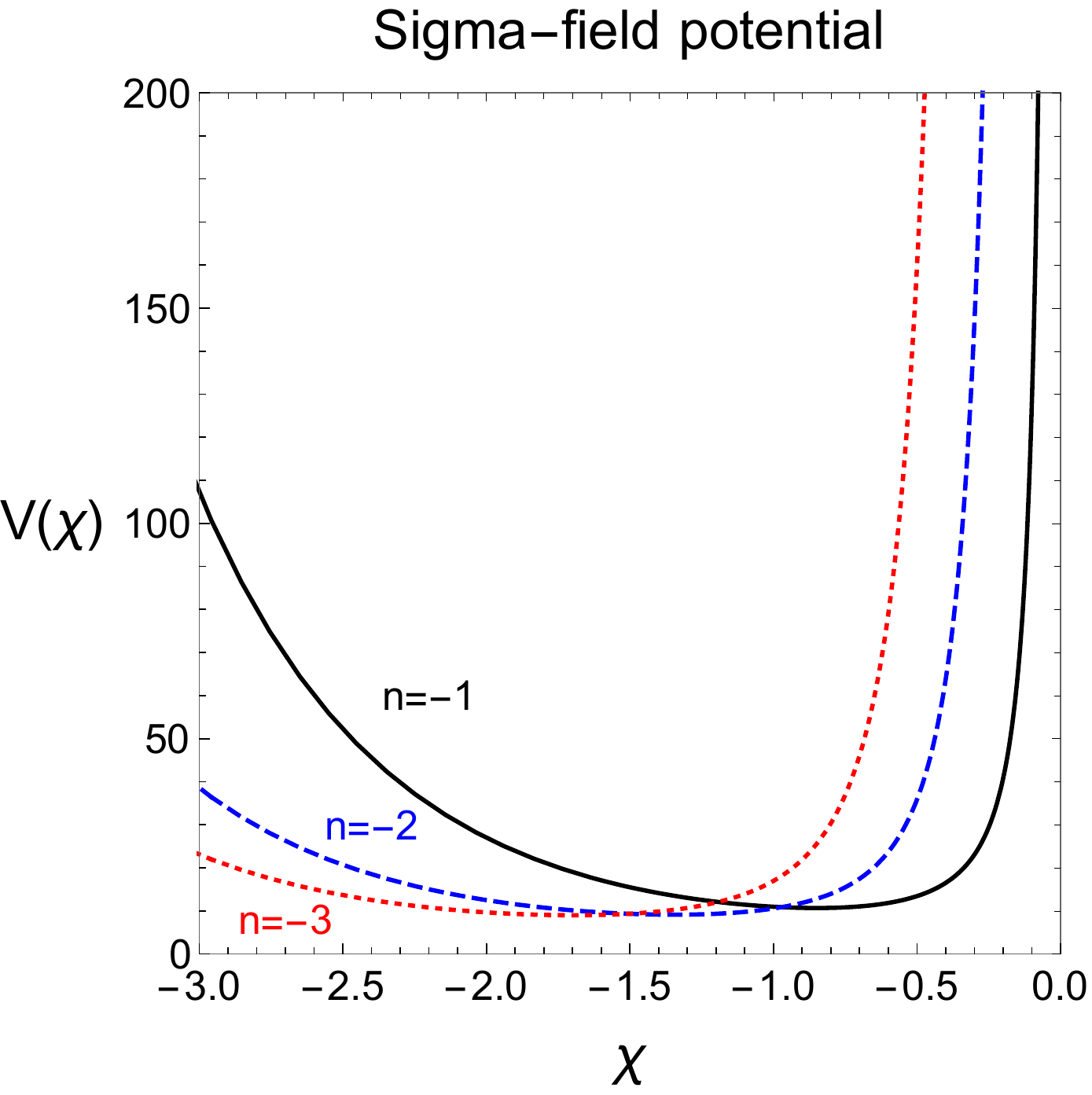}
      \end{center}
  \caption{The sigma field potential for $n=-1, -2, -3$ in black solid, blue dashed, and red dotted lines, respectively. }
  \label{potential}
\end{figure}

For a concrete discussion on dark energy, we take the following form of the potential deviating from  a single exponential, 
\bea
V=V_0\, e^{-c(\chi)\chi},
\eea
with $V_0$ being a constant and $c(\chi)$ being a varying parameter.
For the evolution of the Universe with dark energy, we need to solve the Friedmann equation together with the scalar field equation, given as follows,
\bea
H^2&=& \Big(\frac{\dot a}{a}\Big)^2 = H^2_0 \bigg(\Omega_{m0} \Big(\frac{a_0}{a} \Big)^3 +\Omega_{r0} \Big(\frac{a_0}{a} \Big)^4 +\Omega_{\chi 0}\, \cdot \frac{\rho_\chi}{\rho_{\chi, 0}}     \bigg),  \label{friedeq} \\ 
0&=& {\ddot \chi} + 3H {\dot \chi} + \frac{\partial V}{\partial\chi} \label{deeq}
\eea
where $H_0$ is the Hubble parameter at present, and $\Omega_{i, 0}=\rho_{i,0}/(3M^2_P H^2_0)$ with $i=m, r, \chi$ are the fractions of the energy densities at present for matter, radiation and dark energy, and $\rho_{\chi, 0}$ is the density for dark energy at present.

Then, denoting the equation of state and the energy fraction for $\chi$ by $w$ and $\Omega_\chi$, respectively, and the equation of state for matter by $w_m$, we can determine the attractor behavior for dark energy at late times by recasting eqs.~(\ref{friedeq}) and (\ref{deeq}) into the set of the following equations \cite{quint-review},
\bea
w' &=&(w-1) \Big[3(1+w)-c\sqrt{3(1+w)\Omega_\chi} \Big],  \label{weq}\\
\Omega'_\chi &=& -3(w-w_m)\Omega_\chi (1-\Omega_\chi), \label{omegad} \\
c' &=&-\sqrt{3(1+w)\Omega_\chi}\, (\Gamma-1) c^2 \label{lamd}
\eea
where the prime means the derivative with respect to $\ln a$ with $a$ being the scale factor, and 
\bea
\Gamma= V\, \frac{d^2 V}{d\chi^2}\, \bigg(\frac{d V}{d\chi}\bigg)^{-2}. \label{gamma}
\eea
Here, a constant $c$ leads to $\Gamma=1$. In this case, there is a transition from the early period with $\Omega_\chi=0$, the effective equation of state given by $w_{\rm eff}=w_m$ but undetermined $w$, towards the cosmic acceleration with $\Omega_\chi=1$ and $w=-1+c^2/3$ for $c^2<2$ \cite{quint,quint-review}.
We remark that there was a recent discussion on the lower value of the Hubble constant $H_0$ in quintessence-like models as compared to $\Lambda$CDM \cite{tension}, which would make the tension with the local determinations more serious.

 \begin{figure}
  \begin{center}
    \includegraphics[height=0.42\textwidth]{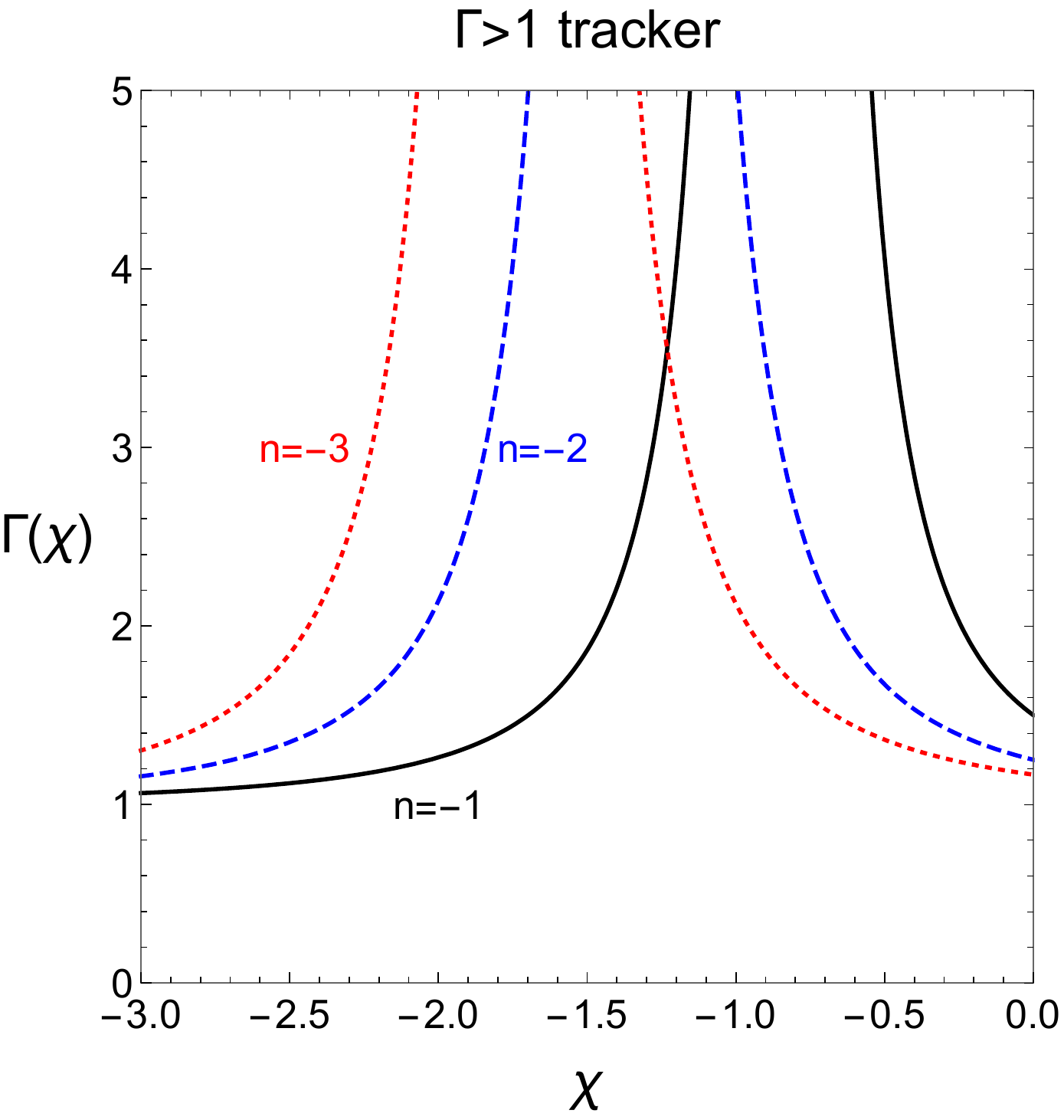}\,\,\,\,
      \includegraphics[height=0.42\textwidth]{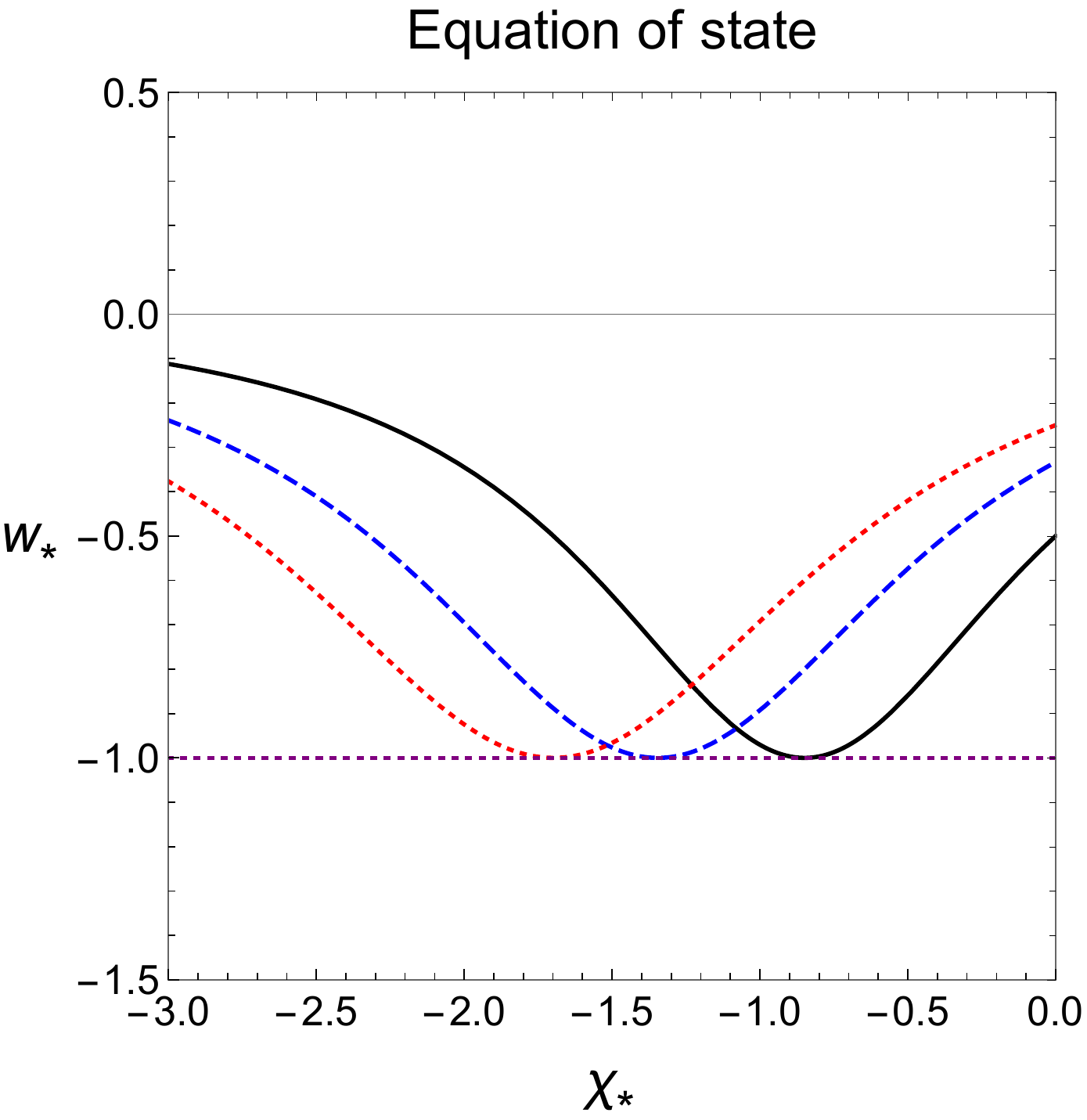}
      \end{center}
  \caption{(Left) The criterion for the tracker condition $\Gamma$ as a function of $\chi$. (Right) The equation of state for the tracker solution as a function of $\chi_*$. $n=-1, -2, -3$ are taken in black solid, blue dashed, and red dotted lines, respectively. }
  \label{tracker}
\end{figure}

In general, for a varying $c$, we need $\Gamma>1$ for $c$ to decrease in time, so the system can evolve into the tracker solution at late times, the epoch of cosmic acceleration. From the potential in eq.~(\ref{pures}), we obtain the $\Gamma$ parameter in eq.~(\ref{gamma}) as
\bea
\Gamma(\chi)=\frac{(n-1)^2+\frac{1}{2}(3n-4)e^{-2\chi/\sqrt{6}} + e^{-4\chi/\sqrt{6}} }{\big(n-1 + e^{-2\chi/\sqrt{6}} \big)^2}, \label{gamma1}
\eea
which depends on the $\chi$ field value. 
In this case, the tracking condition $\Gamma>1$ is satisfied for $n<0$ or $-1<p<0$, independent of $\chi$. That is,  the effective curvature term $R^{p+1}$ with $-1<p<0$ would be appropriate for describing dark energy at present.
In Fig.~\ref{potential}, we depicted the sigma-field potential with arbitrary scales for $n=-1, -2, -3$ (or $p=-\frac{1}{3}, -\frac{1}{5}, -\frac{1}{7}$) in black solid, blue dashed, and red dotted lines, respectively. 
We remark that the fractional power of the higher curvature term, $R^{p+1}$, with $-1<p<0$, could be attributed to the logarithmic running of the Einstein term. For instance, the loop corrections to the Einstein term at the renormalization scale $\mu=R$ \cite{EH-run} take $\alpha_{p} [R-\varepsilon(\ln R) R]\simeq \alpha_{p}\, R^{1-\varepsilon}$ for $|\varepsilon|\ll 1$, in which case we can identify $p=-\varepsilon$.

 \begin{figure}
  \begin{center}
    \includegraphics[height=0.42\textwidth]{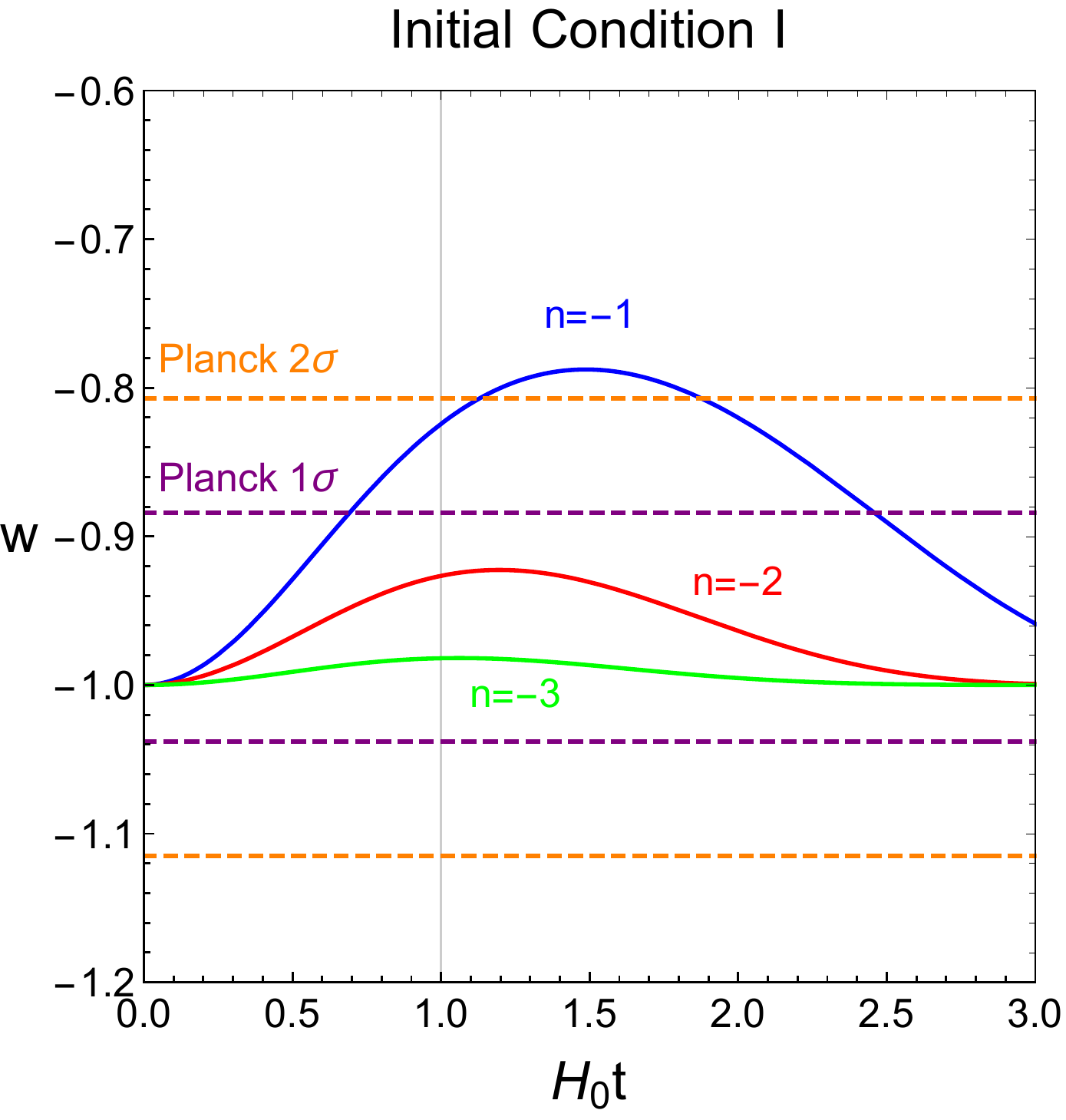}\,\,\,\,
      \includegraphics[height=0.42\textwidth]{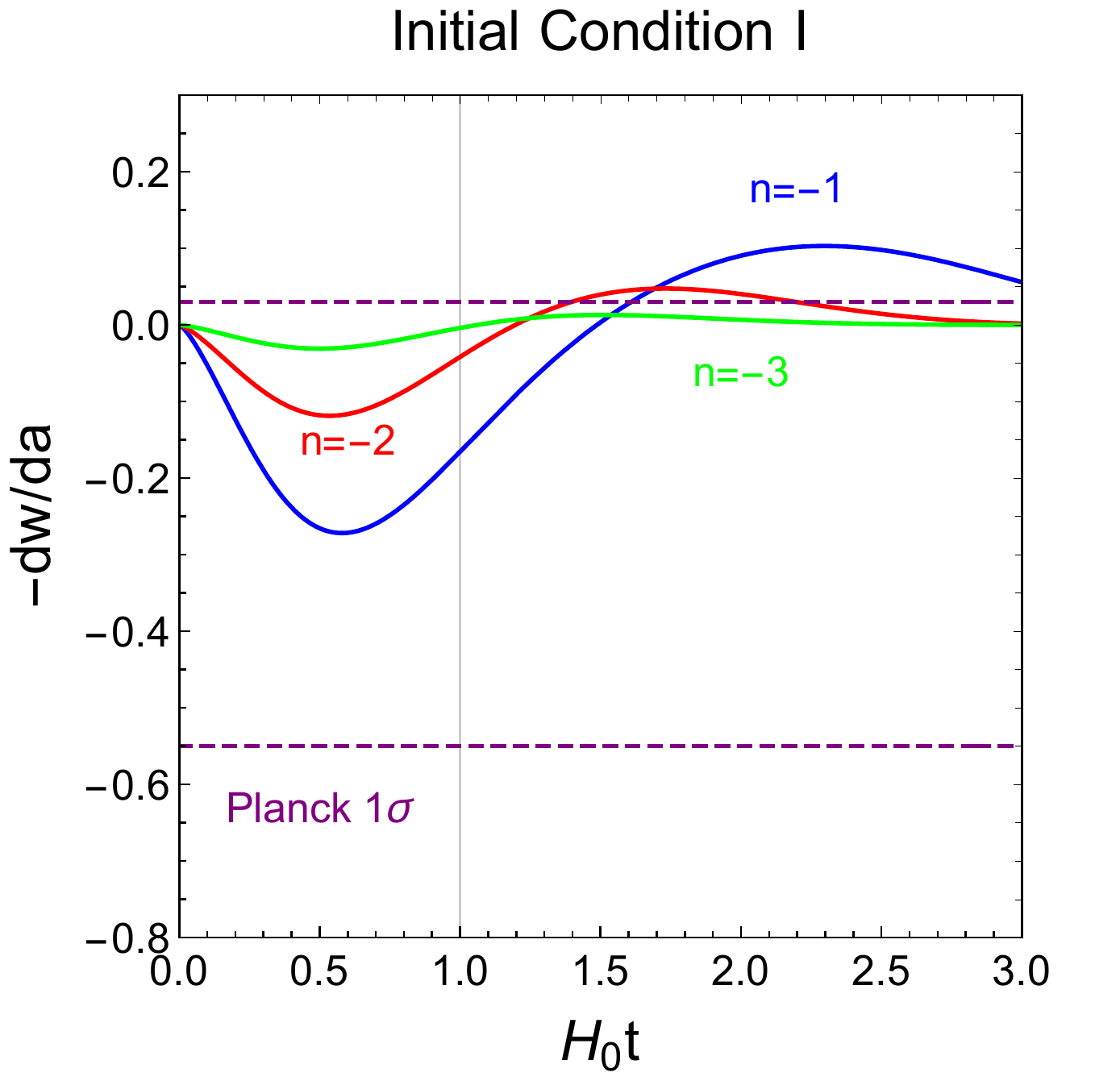}\vspace{0.4cm} \\
        \includegraphics[height=0.42\textwidth]{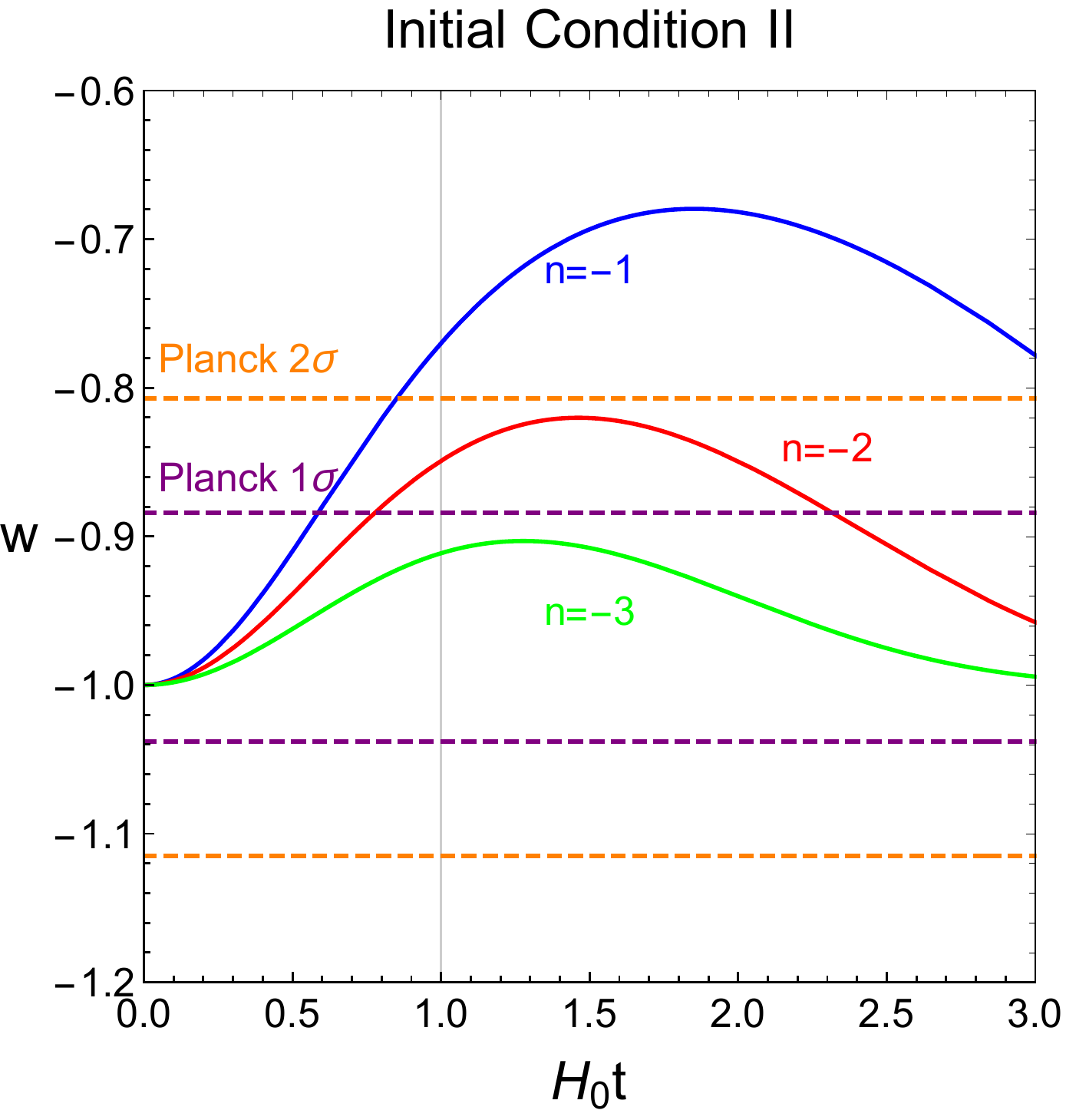}\,\,\,\,
      \includegraphics[height=0.42\textwidth]{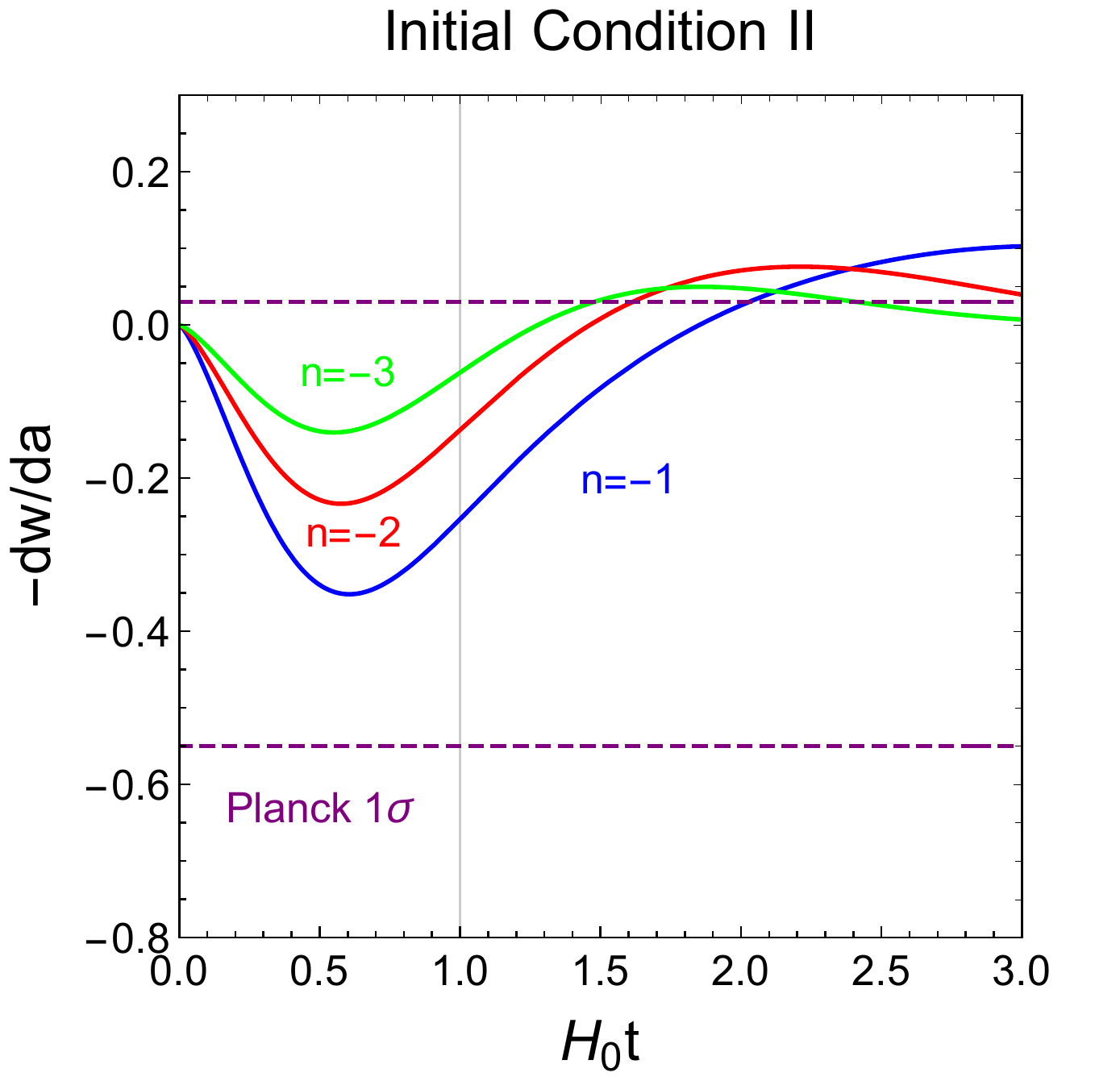} 
      \end{center}
  \caption{Equation of state $w$ and variation of equation of state, $-dw/da$, as functions of $H_0 t$.  $n=-1, -2, -3$ are taken in blue, red and green solid lines, respectively. We chose the initial conditions, $\phi_i=-2M_P, {\dot \phi}_i=0$ for the upper panel, and $\phi_i=-2.5M_P, {\dot \phi}_i=0$ for the lower panel. The vertical gray line is when $a=1$ so $H_0 t\simeq1$. }
  \label{exact}
\end{figure}

For the tracker solution with  $\Gamma>1$, the energy fraction $\Omega_\chi$ and the constant equation of state $\omega_*$ are given \cite{quint-review} by
\bea
\Omega_\chi&=&\frac{3(1+w_*)}{c^2(\chi_*)}, \\
w_*&=& \frac{w_m-2(\Gamma_*-1)}{2\Gamma_*-1}  \label{eos}
\eea
where $\Gamma_*=\Gamma(\chi_*)$ with $\chi_*$ being the field value in the tracker solution.
Here, the latter is obtained  from eqs.~(\ref{omegad}) and (\ref{lamd}) for $\Omega_\chi\ll 1$ during matter domination. 
Then, for $w_m=0$, from eqs.~(\ref{eos}) and (\ref{gamma1}) at $\chi=\chi_*$,  we get the equation of state for dark energy as
\bea
w_* = \frac{n\, e^{-2\chi_*/\sqrt{6}} }{(n-1)^2+(n-2)e^{-2\chi_*/\sqrt{6}} +e^{-4\chi_*/\sqrt{6}}  }
\eea
where we took the $\chi$ field value at $\chi=\chi_*$ at which our approximation with $\Omega_\chi\ll 1$ is valid. 
We note that at $e^{2\chi_m/\sqrt{6}}=\frac{1}{1-n}<1$ with $n<0$, the equation of state becomes $w=-1$ whereas the $\Gamma$ parameter blows up because $\frac{dV}{d\chi}=0$ at $\chi=\chi_m$. Therefore, in order to satisfy the constraints on dark energy \cite{planck2018}, we need to choose $w_*$ close to  the value at $\chi_*=\chi_m$, so the equation of state at present depends on the initial condition in our model.

We drew the tracker condition for $\Gamma$ as a function of $\chi$ on the left plot of Fig.~\ref{tracker} and the equation of state for the tracker on the right plot of Fig.~\ref{tracker}. Here, we took $n=-1,-2,-3$  in black solid, blue dashed, and red dotted lines, respectively, as in Fig.~\ref{potential}. The smaller $n$, the $\chi$ field value required for $w_*=-1$ moves towards a negative value. 

In Fig.~\ref{exact}, we depicted the numerical solutions for the equation of state $w$ on left and the variation of the equation of state, $-dw/da$, on right, in each panel. They are shown as functions of $H_0 t$.  We took $n=-1,-2,-3$ in 
blue, red and green solid lines, respectively, and chose the initial conditions, $\phi_i=-2M_P, {\dot \phi}_i=0$ for the upper panel (Initial Condition I), and $\phi_i=-2.5M_P, {\dot \phi}_i=0$ for the lower panel (Initial Condition II). We also took the initial value of the scale factor $a_i=3\times 10^{-6}$ at radiation domination for the numerical analysis, but we can always choose a larger initial value of the scalar field for a smaller value of $a_i$. 

As compared to the parametrization of the time-varying equation of state,
\bea
w(a) =w_0 + (1-a) w_a
\eea
where $w_0=w(a=1)$ and $w_a=-\frac{dw}{da}(a=1)$,  the constraints on $w_0, w_a$ from Planck $+$ SNe $+$ BAO are given by $w_0=-0.961\pm 0.077$ and $w_a=-0.28^{+0.31}_{-0.27}$ \cite{planck2018}. Therefore, in Fig.~\ref{exact}, we imposed those constraints at $1\sigma$ and $2\sigma$ levels within the purple and orange dashed lines, respectively. 
As a result, we find that the cases with $n=-2, -3$ are consistent with dark energy constraints within $1\sigma$, but $n=-1$ is in tension with the data at $2\sigma$ or more.  We find that as the initial value of the $\chi$ field gets smaller, the results are more consistent with the data.  Furthermore, as $n$ becomes more negative, that is, $|p|$ gets even smaller, the  equation of state gets closer to $w=-1$, as can be seen from the tendency in the left plots in the upper and lower panels in  Fig.~\ref{exact}.

\subsection{The effect of the Higgs field couplings}

In the presence of the Higgs field, we need to consider the effect of the Higgs field for the effective sigma-field potential. 
We recall the total scalar potential for a general $n$ in Einstein frame,
\bea
V(\sigma,h)&=& \frac{1}{\big(1-\frac{1}{6}h^2 -\frac{1}{6}\sigma^2\big)^2}\bigg[ \frac{\kappa_n}{4}\,(\sigma+\sqrt{6})^{4(1-n)}\bigg(-\sigma(\sigma+\sqrt{6})-3\Big(\xi+\frac{1}{6} \Big)h^2\bigg)^{2n}  \nonumber \\
&&+\frac{1}{4}\lambda (h^2-v^2)^2\bigg]
\eea 
where we have included the Higgs mass term and the cosmological constant in Jordan frame.
Then, for the canonical sigma field in eq.~(\ref{inflaton}) and taking $|\sigma^2-6|\gg h^2$, the Higgs interactions modify the potential for dark energy to
\bea
V(\chi,h) &=& V_0(\chi)\cdot \bigg(1-\frac{1}{4} \Big(\xi+\frac{1}{6} \Big)h^2\,\cdot\frac{e^{2\chi/\sqrt{6}} (1+e^{-2\chi/\sqrt{6}})^2}{(1-e^{-2\chi/\sqrt{6}})} \bigg)^{2n} \nonumber \\
&&+\frac{9}{16}\lambda \,e^{-4\chi/\sqrt{6}} \Big(1+e^{2\chi/\sqrt{6}}\Big)^4 (h^2-v^2)^2. 
\eea

Consequently, after electroweak symmetry breaking with $\langle h\rangle=v$, the full scalar potential is reduced to eq.~(\ref{pures}), up to a small correction proportional to $(\xi+1/6)v^2/M^2_P$.  On the other hand, before electroweak symmetry breaking, the effective Higgs quartic coupling can run with the sigma field, 
\bea
\lambda_{\rm eff} =\frac{9}{4}\,\lambda\, e^{-4\chi/\sqrt{6}} \Big(1+e^{2\chi/\sqrt{6}}\Big)^4.
\eea
In our case, as discussed in the previous subsection, the minimum value of the equation of state, $w=-1$,  appears for $\chi<0$, so the effective Higgs quartic coupling can get larger than in the SM \cite{vsb,swamp}.  But, from eq.~(\ref{yukawa}) in Appendix B, the effective top Yukawa coupling also scales with the sigma field by
\bea
y_{\rm eff}&=&\frac{y_t}{\big(1-\frac{1}{6}h^2-\frac{1}{6}\sigma^2\big)^{1/2}} \nonumber  \\
&=& 2y_t\, e^{-\chi/\sqrt{6}} \Big(1+e^{2\chi/\sqrt{6}}\Big).
\eea
Therefore, the effective top Yukawa coupling gets stronger for $\chi<0$, leading to the field-dependent beta function for the Higgs quartic coupling, which grows negatively with $-y^4_{\rm eff}\sim -y^4_t\, e^{-4\chi/\sqrt{6}}$.  Thus, the result is in contract to the case with the overall quintessence coupling, motivated by the dS swampland conjecture \cite{swamp}, in which case the potential is given by $V(\chi)=e^{-c\chi}\big(\frac{1}{4}\lambda(h^2-v^2)^2+\Lambda\big)$, with $\Lambda$ being the constant parameter for dark energy  and the top Yukawa coupling does not run with the quintessence field.

\section{Conclusions}
We have presented the general linear sigma models with conformal invariance as the UV completion of the Higgs inflation and made a concrete realization of them in the context with general higher curvature terms beyond Einstein gravity. 
Thus, we have identified the dual-scalar theory for the higher curvature terms, $R^{p+1}$ with $p>0$, as the UV complete models for the Higgs inflation. 

The successful inflation singles out a particular linear sigma model coming from the $R^2$ term among a class of general linear sigma models. In the basis where the Higgs field and the dual scalar for the $R^2$ term satisfy conformal invariance, we identified the effective inflaton potential after the Higgs field is integrated out during inflation, and compared the inflationary predictions to  those in the literature. 

We have also shown that as the outcome of the general dual-scalar formulation, the higher curvature terms, $R^{p+1}$, with $-1<p<0$, which could be originated from the logarithmic running of the Einstein term at loops, can provide dark energy with the tracker behavior  at late times as in quintessence models. In this case, we found that the model predictions for the time-varying equation of state for dark energy can be consistent with Planck, SNe and BAO data within $1\sigma$.  More precise measurements of the equation of state for dark energy in the future experiments could narrow down the upper bound on $p$ in these models. We also discussed the implication of the Higgs-sigma interactions for the running quartic coupling in the early Universe and the vacuum stability problem in the SM.

\acknowledgments

We would like to thank Yohei Ema for helpful communications on the subject. 
The work is supported in part by Basic Science Research Program through the National Research Foundation of Korea (NRF) funded by the Ministry of Education, Science and Technology (NRF-2019R1A2C2003738 and NRF-2021R1A4A2001897). 
The work of AGM is supported in part by the Chung-Ang University Young Scientist Scholarship in 2019.

\appendix
\section{ Relation to  the induced gravity}

We discuss the relation between Higgs inflation with $R^2$ term and  the induced gravity model.
Starting from the scalar-dual Lagrangian in eq.~(\ref{r2-dual}),  we can make a field redefinition by
\bea
\sigma=1+\xi{\hat\phi}^2_i +4\alpha {\hat\chi}. 
\eea
Then, the resulting Lagrangian is
\bea
\frac{{\cal L}_{R2}}{\sqrt{-{\hat g}}} = -\frac{1}{2} \sigma {\hat R} - \frac{1}{16\alpha}\, (\sigma-1-\xi {\hat\phi}^2_i)^2 + \frac{1}{2} (\partial_\mu{\hat\phi}_i)^2 -\frac{\lambda}{4} ({\hat\phi}^2_i)^2. 
\eea
This belongs to the category of the induced gravity models for unitarizing Higgs inflation, as discussed in Ref.~\cite{sigma}. But, in this case, the conformal symmetry is not manifest in the derivative terms.
With a Weyl rescaling by ${\hat g}_{\mu\nu}=g_{\mu\nu}/\sigma$, we can get the Einstein-frame Lagrangian,
\bea
\frac{{\cal L}_{R2}}{\sqrt{-{g}}}=  -\frac{1}{2}  R + \frac{3}{4\sigma^2} \,(\partial_\mu\sigma)^2 -\frac{1}{16\alpha \sigma^2}\,  (\sigma-1-\xi {\hat\phi}^2_i)^2 + \frac{1}{2\sigma} (\partial_\mu{\hat\phi}_i)^2 -\frac{\lambda}{4\sigma^2} ({\hat\phi}^2_i)^2.
\eea
We note that there are a lot of possible field redefinitions for bringing the Higgs non-minimal coupling to the sigma field potential and unitarizing Higgs inflation as far as the couplings in the potential are perturbative. In the above case, the full Higgs quartic coupling is given by
\bea
\lambda_{\rm full}= \lambda + \frac{\xi^2}{4\alpha}.
\eea
Thus,  as far as as $\xi^2/(4\alpha)\lesssim 1$, there is no perturbativity problem.

\section{From conformal frame to Einstein frame}

We note that the original metric ${\hat g}_{\mu\nu}$, for which the Higgs non-minimal coupling in eq.~(\ref{Higgsinf}) is introduced, is related to the Einstein metric $g_{E,\mu\nu}$ by
\bea
{\hat g}_{\mu\nu} = \Omega^{-2} g_{\mu\nu} =\Omega^{-2}_{\rm eff}\, g_{E,\mu\nu}
\eea
where the effective conformal factor for the Weyl rescaling is given by
\bea
\Omega^{2}_{\rm eff}= \Omega^{2}\Omega^{\prime 2}  =  \frac{1-\frac{1}{6} h^2-\frac{1}{6}\sigma^2}{\big(1+\frac{\sigma}{\sqrt{6}}\big)^2}.
\eea
On the other hand, the original Higgs fields ${\hat\phi}_i$ defined in the  original frame with ${\hat g}_{\mu\nu}$ are also related to those $\phi_i$ defined in the Einstein frame with $g_{E,\mu\nu}$ by
\bea
{\hat\phi}_i=\Omega \phi_i = \Big(1+\frac{\sigma}{\sqrt{6}}\Big)^{-1}\, \phi_i.
\eea
As a result, after a rescaling of the fermion field $\hat\psi$ in the original frame with  ${\hat g}_{\mu\nu}$ by
\bea
{\hat\psi} = \Omega^{3/2}_{\rm eff}  \psi,
\eea
under which the kinetic term for the fermion field remains unchanged,
we can identify the Yukawa coupling for the redefined Higgs by
\bea
{\cal L}_Y &=&\sqrt{-{\hat g}} \Big(-y_f  {\hat H} \overline {{\hat\psi}}_L {\hat\psi}_R +{\rm h.c.} \Big) \nonumber \\
 &=&\sqrt{-g} \Big(-y_f \Omega\, \Omega^{-1}_{\rm eff} H {\bar\psi}_L \psi_R +{\rm h.c.}\Big) \nonumber \\
 &=&\sqrt{-g} \Big(-y_f (\Omega^\prime)^{-1} H {\bar\psi}_L \psi_R +{\rm h.c.}\Big) \label{yukawa}
\eea
with
\bea
(\Omega^\prime)^{-1}= \Big(1-\frac{1}{6} h^2-\frac{1}{6}\sigma^2\Big)^{-1/2}.
\eea


\end{document}